\documentclass[11pt,a4paper]{article}
\usepackage{jheppub,amsmath,  amssymb,slashed,url,bm,textgreek,upgreek}
\usepackage{graphicx}
\usepackage{epstopdf}
\def\t{{ \sf t}} 
\def\tt{{\mathfrak t}}

\def\X{{\mathcal X}}

\def\be{\begin{equation}}
\def\ee{\end{equation}}

\def\X{{\eusm X}}
\def\Re{{\mathrm{Re}}}
\def\Im{{\mathrm{Im}}}
\def\hat{\widehat}

\def\tilde{\widetilde}

\def\RP{{\Bbb{RP}}}

\def\D{{\mathcal D}}

\def\Bbb{\mathbb}

\def\d{{\mathrm d}}

\def\R{{\mathbb R}}
\def\C{{\mathbb C}}

\def\D{{\mathcal D}}
\def\[{\bigl [}

\def\]{\bigr ]}

\def\CP{{\mathbb{CP}}}

\def\T{{\mathcal T}}

\def\Z{{\mathbb Z}}

\def\t{\widetilde }

\def\B{{\mathcal B}}

\def\M{{\mathcal M}}

\def\W{{\mathcal W}}

\def\vol{{\mathrm{vol}}}

\def\tilde{\widetilde}

\def\bar{\overline}

\font\teneurm=eurm10 \font\seveneurm=eurm7  \font\fiveeurm=eurm5
\newfam\eurmfam
\textfont\eurmfam=\teneurm \scriptfont\eurmfam=\seveneurm
\scriptscriptfont\eurmfam=\fiveeurm

\font\teneusm=eusm10 \font\seveneusm=eusm7 \font\fiveeusm=eusm5
\newfam\eusmfam
\textfont\eusmfam=\teneusm \scriptfont\eusmfam=\seveneusm
\scriptscriptfont\eusmfam=\fiveeusm
\def\eusm#1{{\fam\eusmfam\relax#1}}
\font\tencmmib=cmmib10 \skewchar\tencmmib='177
\font\sevencmmib=cmmib7 \skewchar\sevencmmib='177
\font\fivecmmib=cmmib5 \skewchar\fivecmmib='177
\newfam\cmmibfam
\textfont\cmmibfam=\tencmmib \scriptfont\cmmibfam=\sevencmmib
\scriptscriptfont\cmmibfam=\fivecmmib

\title{The Feynman $i\varepsilon$ in String Theory}

 \author{Edward Witten}
\affiliation{School of Natural Sciences, Institute for Advanced Study,\\ 1 Einstein Drive, Princeton, NJ 08540 USA}
\abstract{The Feynman $i\varepsilon$ is an important ingredient in defining perturbative scattering amplitudes in
field theory.  Here we describe its analog in string theory.  Roughly one takes the string worldsheet
to have Lorentz signature when a string is going on-shell although it has Euclidean signature generically.  (This article is
based on a talk presented at the Zuminofest in Berkeley, California, May 2-4, 2013.)
}

\begin{document}\maketitle

\section{Introduction}\label{intro} 

In signature $-++\dots+$, the Feynman propagator for a scalar field of mass $m$ and momentum $p$ is
\begin{equation}\label{feynmann}\frac{-i}{p^2+m^2-i\varepsilon},\end{equation}
where $\varepsilon$ is a small positive quantity that should be taken to zero at the end of a calculation.   This
prescription for treating the singularity of the propagator at $p^2+m^2=0$ ensures causality in the scattering of wave packets
at tree level, and is essential in getting correct and physically sensible loop amplitudes.  

The purpose of the present
paper is to explain what the Feynman $i\varepsilon$ means in the context of string perturbation theory.  Actually,
in string field theory -- whether light cone string field theory or covariant string field theory -- the answer is clear: one literally includes a $-i\varepsilon$
in the denominator of the string propagator, shifting $1/L_0$ to $1/(L_0-i\varepsilon)$.  Our goal here is more specifically to explain what the $i\varepsilon$
means in covariant approaches to string perturbation theory in which scattering amplitudes are computed by integration over a suitable moduli space of Riemann
surfaces or super Riemann surfaces.   To find an answer, we  merely  translate the procedure one would follow in string field theory into 
the language of integration over moduli space.   In this way, one finds 
that the $i\varepsilon$ amounts to a certain deformation of the integration cycle used in string perturbation theory.  The deformation is
only relevant near infinity, where the string worldsheet degenerates. 

The question of interpreting the $i\varepsilon$ in string theory was raised in \cite{gross} and perhaps elsewhere.  
The basic idea of the 
present paper has been previously described in \cite{berera}, where an explicit example is discussed, and very briefly in \cite{mandelstam}.  It probably
has also been known to other researchers.  
The goal of the present paper is to fill in details, make clear that there is a systematic procedure, and describe  simple consequences.
For early work on 1-loop diagrams
involving questions related to the $i\varepsilon$, see \cite{bo,amano, weis,dph}.

\section{Singularities And Propagators}\label{zepsilon}

\subsection{Basic Tree Amplitudes}\label{sometree}

For orientation, we begin by considering elastic two-body scattering at tree level.
As usual, the four external states are represented by momentum vectors $p_1,\dots,p_4$, constrained by momentum conservation $\sum_i p_i=0$ 
and by mass shell conditions $p_i^2=-m_i^2$, where $m_i$ is the mass of the $i^{th}$ particle.  
One introduces the usual kinematic invariants $s=-(p_1+p_2)^2$, $t=-(p_1+p_3)^2$, $u=-(p_1+p_4)^2$,
obeying
\begin{equation}\label{xelbo} s+t+u=\sum_i m_i^2. \end{equation}

We begin with tachyon scattering in bosonic open string theory. The tachyon mass squared is $m^2=-1/\alpha'$,
so the condition (\ref{xelbo}) is $s+t+u=4m^2=-4/\alpha'$.  The Veneziano amplitude for tachyon elastic scattering is
\begin{equation}\label{venez}A_{\mathrm{V}}(s,t)=\int_0^1\d x\, x^{-\alpha' s -2}(1-x)^{-\alpha' t-2}.   \end{equation}
The integral over $x$ is a special case of an integral over a moduli space of Riemann surfaces in string theory: we integrate over the moduli space that parametrizes a disc with four cyclically ordered marked points on its boundary (namely $0,x,1,\infty$).

The integral (\ref{venez}) converges near $x=0$ and $x=1$ for $s,t<m^2=-1/\alpha'$, which is compatible with (\ref{xelbo}) if $u>-2/\alpha'$.
Thus there is a region in which the integral over moduli space converges, but this region does not include the region of physical $s$- or $t$-channel scattering.   The values $s=-1/\alpha'$ and $t=-1/\alpha'$ above which the integral diverges are simply the thresholds for the lightest
physical state (the tachyon) in the $s$- or $t$-channel.  If we take $s$ and $t$ to be complex, the condition for the integral to converge is $\Re\,s,\,\Re\, t
<-1/\alpha'$.  Here and in all similar statements below, there can be conditional convergence when one of these inequalities becomes an equality,
for instance if  $\Re\,s=-1/\alpha'$ and $\Im \,s \not=0$.

\begin{figure}
 \begin{center}
   \includegraphics[width=3in]{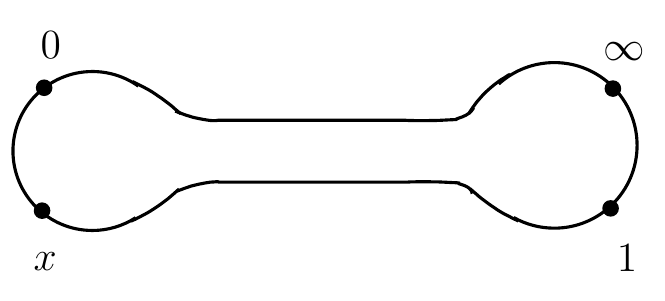}
 \end{center}
\caption{\small A disc with marked points $0,x,1,\infty$ on its boundary is conformally equivalent, for $x\to 0$, to the Riemann surface with boundary depicted here, which
describes propagation of an open string through a proper time of order $|\log x|$. }
 \label{equivo}
\end{figure}
The Veneziano amplitude actually has a pole at threshold, that is at $s=-1/\alpha'$ or $t=-1/\alpha'$.  (Of course, it also has infinitely many poles
above threshold.) Like all singularities in string theory, these poles are infrared singularities associated to an on-shell string state -- in this case, a tachyon
propagating in the $s$-channel or $t$-channel. In fact,  as $x\to 0$, the string worldsheet is conformally equivalent (fig. \ref{equivo})
to one that describes propagation of a string through a large proper time, leading to an on-shell pole.  Thus the pole of the Veneziano amplitude
at, say, $s=-1/\alpha'$ is analogous
to the pole of the Feynman propagator $-i/(p^2+m^2-i\varepsilon)$ at $p^2+m^2-i\varepsilon=0$.  (As usually defined, the Veneziano amplitude
does not have a shift in the pole by $i\varepsilon$, and this is part of what we will grapple with.  The factor of $-i$ in the numerator is also not immediately apparent in eqn. (\ref{venez}).  To find
it, multiply eqn. (\ref{venez}) and all other formulas in this paper for tree amplitudes by an overall factor of $i$, as is appropriate for  Lorentz signature tree amplitudes.  Then when one factorizes
an amplitude at a pole as the product of two subamplitudes times a propagator, the propagator has a factor of $-i$.)

For bosonic closed string tachyons, which have $m^2=-4/\alpha'$, the four particle tree amplitude is the Virasoro-Shapiro amplitude
\begin{equation}\label{vis}A_{\mathrm{ViS}}=\int_\C\d^2 z \,|z|^{-\frac{1}{2}\alpha's-4}|1-z|^{-\frac{1}{2}\alpha't -4}.\end{equation}
The integral is over the moduli space that parametrizes a genus 0 Riemann surface (the Riemann sphere) with four marked points (namely
$0,1,\infty$, and $z$).   This amplitude has poles in all channels, beginning at $s,t,$ or $u$ equal to $m^2=-4/\alpha'$, and
convergence at $z=0,1,$ and $\infty$ requires that $\Re\,s,\,\Re\, t$, and $\Re\,u$ should all be below threshold, 
that is less than $m^2$.  This is consistent with (\ref{xelbo}), which in this case gives $s+t+u=4m^2=-16/\alpha'$,
but only because the external
states are tachyons, with $m^2<0$. 

Tree-level elastic scattering of massless or massive string states -- in bosonic string theory or in any of the superstring theories -- is given by similar integral
expressions, but with shifted exponents controlling the behavior near $x=0,1$ for open strings or near $z=0,1,\infty$ for closed strings.  For
open strings, the condition for the integral to converge is always that $\Re\, s$ and $\Re\, t$ are below the threshold set by the lowest mass
pole (or resonance) in the given channel.  For closed strings, convergence requires $\Re\,s$, $\Re\,t$, and $\Re\,u$ to be all  below threshold.
For massive external closed string states, these conditions are  incompatible with (\ref{xelbo}) 
and the integral never converges.  For massless external closed string states, the same statement holds except that the integral is conditionally convergent
if $s,t,$ and $u$ are all imaginary and nonzero.

Similar  issues arise in $n$-particle tree amplitudes with $n>4$.  For example, the tree-level $n$-point function for bosonic string
tachyons is defined formally
by the integral
\begin{equation}\label{gurf}A_{n,\mathrm{open}}=\frac{1}{\vol\, SL_2(\R)}\int_S \d x_1\dots \d x_n\prod_{i<j} |x_i-x_j|^{2\alpha' p_i\cdot p_j} .\end{equation}
Here the set $S$ parametrizes points $x_1,\dots,x_n\in \R\cup\infty =\RP^1$ in a given cyclic order (say $x_1<x_2<\dots<x_n$); also, $1/\vol\, SL_2(\R)$
is shorthand for a standard procedure of dividing out by the action of $SL_2(\R)$.  For general $n$, it is unclear that there any choices of external momenta
for which this integral converges.  Tree-level scattering of closed string tachyons is given by a similar formula in which the $x_i$ are replaced by
complex variables:
\begin{equation}\label{wurf}A_{n,\mathrm{closed}}=\frac{1}{\vol \,SL_2(\C)}\int \d^2z_1\dots \d ^2z_n\prod_{i<j} |z_j-z_j|^{\alpha' p_i\cdot p_j} .\end{equation}
The points $z_i$ range independently over the Riemann sphere, modulo the action of $SL_2(\C)$.
Convergence is only harder to come by, since there are more channels in which an infrared divergence might arise.

\subsection{Direct Approaches To Analytic Continuation}\label{naive}

Clearly, the region in which the integral over the moduli space of Riemann surfaces converges is often empty and never suffices to describe physical string theory
scattering processes.   In the specific case of the Veneziano amplitude, there is a safe region ($s,t<-1/\alpha'$) in which the integral over moduli space
converges, and one can consider the integral to be defined by analytic continuation from there.   A general string theory amplitude does not have such a safe
region, so one needs a better approach.

\subsubsection{Splitting The Integral}\label{splitting}
One slightly {\it ad hoc} approach is to split the integral over moduli space as a sum of several terms each of which has better convergence than the full
integral.  Then one analytically continues each piece separately.  For example, one could split the Veneziano amplitude as a sum of two
pieces
\begin{equation}\label{zork}A_{\mathrm V}(s,t)=\int_0^{1/2}\d x\, x^{-\alpha' s -2}(1-x)^{-\alpha' t-2}+\int_{1/2}^1\d x\, x^{-\alpha' s -2}(1-x)^{-\alpha' t-2},\end{equation}
where the first integral converges for $\Re\, s<-1/\alpha'$ and the second for $\Re\,t<-1/\alpha'$.   Similarly, the integral (\ref{vis}) defining the Virasoro-Shapiro
amplitude could be split as a sum of integrals over three regions, one  centered near $z=0$, one near $z=1$, and one near $z=\infty$, such that each
integral would converge given a suitable condition on just one of $s,t$, or $u$.   Each of the three integrals would define an analytic function in its own region
of convergence, and after analytically continuing these functions and adding them up, one would get the full Virasoro-Shapiro amplitude.  

It seems very likely that this program can be carried out for general tree amplitudes, but the procedure seems a little unphysical since the individual pieces
are unnatural and have bad high energy behavior.  (For example, the large $s$, fixed $t$ behavior  of either of the two terms on the right hand side of (\ref{zork})
is dominated by the behavior near $x=1/2$.)    There may be some real difficulty in this program for loop amplitudes (though it has been an ingredient
in some approaches at 1-loop level \cite{weis,dph}).   This is because loop amplitudes are
complicated multi-branched analytic functions.  If one chops up a loop amplitude as a sum of pieces,
each of which converges in a different region, then in analytically
continuing everything back to the physical region, it might be hard
to put all the pieces together properly and land on the right sheet
of the complex amplitude.  Therefore, we will look for a different approach.

\subsubsection{Pochhammer Contours}\label{pochhammer}

\begin{figure}
 \begin{center}
   \includegraphics[width=3in]{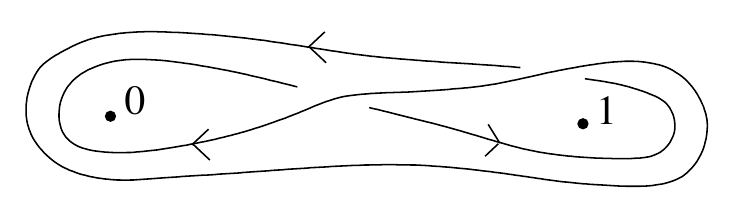}
 \end{center}
\caption{\small The Pochhammer contour is a closed contour that winds back and forth around the branch points at 0 and 1 in such a way that the form $\omega=\d x\,x^{-\alpha' s -2}(1-x)^{-\alpha' t-2}$ is
single-valued. }
 \label{pock}
\end{figure}
A more sophisticated approach to tree amplitudes involves Pochhammer contours \cite{Pock,hsh}.  The first step is to analytically continue the integrand
in the worldsheet path integral.  For example, in the case of the Veneziano amplitude, we view $x$ as a complex variable and we view the differential
form $\omega = \d x \, x^{-\alpha' s -2}(1-x)^{-\alpha' t-2}$ as a (multi-valued) holomorphic 1-form on the complex $x$-plane, with branch points at $0,1$, and $\infty$.  
The Pochhammer
contour 
is a closed contour $\gamma$ in 
 $\C\backslash \{0,1\}$ (that is, in the complex $x$-plane with the branch points removed) on which the form $\omega$ is single-valued.
Such a contour can be found (fig. \ref{pock}) by looping around the branch points at 0 and 1 in a suitable fashion.  Since $\gamma$ is compact,
the integral $\oint_\gamma\omega$ converges for all $s,t$ and defines an entire function of $s$ and $t$.  The Veneziano amplitude
can be expressed in terms of this function:
\begin{equation}\label{gulf} A_{\mathrm V}=\frac{\oint_\gamma\omega}{(1-e^{-2\pi i \alpha' s})(1-e^{-2\pi i \alpha' t})}  .\end{equation}
To prove this formula, one uses the fact that because of the way $\gamma$ zigzags between 0 and 1, the integral $\oint_\gamma\omega$ is equivalent
to the sum of 4 copies of $\int_0^1\omega$, except that the 4 copies must be evaluated on different branches of the multi-valued form $\omega$, and 2 of
them contribute with a minus sign because of a reversed orientation.  The sum over the 4 copies gives 
 $\oint_\gamma\omega=(1-e^{-2\pi i \alpha' s})(1-e^{-2\pi i \alpha' t})\int_0^1\omega$, provided the overall branch of the integrand is properly chosen in each integral.
This leads to (\ref{gulf}),  which (as a formula for the Euler beta function) goes back to \cite{Pock}.
For a generalization of this for a five-particle open-string tree amplitude, see \cite{hsh}.  In principle, there is a generalization for any number of particles,
though the details probably become complicated.

This procedure can also be generalized to closed-string tree amplitudes.   Again  the first step is analytic continuation of the integrand.  We analytically continue
from the complex $z$-plane to $\C\times \C$ by
regarding $z$ and its complex conjugate $\bar z$ as independent complex variables, say $z$ and $\t z$.  (Equivalently, we regard the real and imaginary parts
of $z$ as independent complex variables.)  The form which must be integrated
to define the Virasoro-Shapiro amplitude is then
\begin{equation}\label{zulf} \varpi=\d \t z \,\d z\, (\t z z)^{-\frac{1}{4}\alpha' s-2}((1-\t z)(1-z))^{-\frac{1}{4}\alpha't
-2}.\end{equation}
We view $\varpi$, roughly, as a holomorphic 2-form on a copy of $\C^2$, parametrized by $\t z $ and $z$; to be more precise, $\varpi$ is a holomorphic
but multi-valued 2-form on the complement of a certain branch locus $\Delta$.  The analog of a Pochhammer
contour is a 2-cycle $\lambda\subset \C^2\backslash\Delta$ on which $\varpi$ is single-valued.  To construct such a cycle, one observes that $\varpi$
is the product of a 1-form on the $\t z$-plane and a 1-form on the $z$-plane:
\begin{equation}\label{ozulf} \varpi= \d \t z\,\t z^{-\frac{1}{4}\alpha' s-2}(1-\t z)^{-\frac{1}{4}\alpha't
-2}\cdot\d z\,  z^{-\frac{1}{4}\alpha' s-2}(1-z)^{-\frac{1}{4}\alpha't
-2}.\end{equation} (This factorization results from the fact that all modes of a  bosonic string
 in $\R^{26}$, except for the zero-modes which describe the center of mass
motion of the string,  can be decomposed as sums of holomorphic and antiholomorphic modes.) 
Hence we can take $\lambda$ to be a product of a Pochhammer contour in the $z$-plane and a Pochhammer contour in the
$\t z$-plane.  The Virasoro-Shapiro amplitude can be written as an integral over this cycle, with a prefactor that is somewhat analogous to the one
in (\ref{gulf}).  This fact is related to the KLT formulas \cite{KLT} that express closed-string tree amplitudes in terms of products of open-string tree amplitudes.

Although this approach to tree amplitudes is very elegant, it has a subtle drawback: it does not naturally incorporate the Feynman $i\varepsilon$.  Since the
Pochhammer contour is compact, integration over this contour does not generate any singularities.
Accordingly, the poles of the Veneziano amplitude
are all contained in the explicit prefactor in eqn. (\ref{gulf}) and there is no particular reason to shift these poles by $i\varepsilon$.
At tree level, this may not seem critical as one may add the $i\varepsilon$ by hand.   However, at the loop level, the $i\varepsilon$
is crucial input to define the correct amplitudes and cannot be added as an afterthought. 
Loop amplitudes have very complicated analytic behavior and it does not appear realistic to capture their singularities by an explicit prefactor.
 Therefore, it does not seem likely, at least to the author, that loop amplitudes can be defined by integration over compact
 cycles analogous to the Pochhammer contours that work for tree amplitudes.    

We will follow a different approach that is much more closely related to the way that the Feynman $i\varepsilon$ appears in ordinary
field theory.  However, some steps will be the same as above: we analytically continue from a real integration variable
$x$ to a complex one (or from a complex integration variable $z$ to a pair of complex variables $z,\t z$) and  modify the naive integration
cycle in a way that involves  extending it into the complex domain.

\subsection{Continuation From Euclidean Signature}\label{euclidean}

In field theory, one approach to avoiding any problems associated with the pole of the Feynman propagator is to begin in Euclidean
signature with a correlation function of local operators $\langle \phi_1(x_1)\dots \phi_n(x_n)\rangle$.  In Euclidean signature, the propagator
is $1/(p^2+m^2)$ and there are not any serious problems associated with the zero of the denominator.  Then one analytically continues to Lorentz
signature and extracts the $S$-matrix via the LSZ formula.  There are a few issues to consider. It may be
hard to understand the analytic continuation. Also, the off-shell
continuation that one has to make is not gauge-invariant in the
case of charged particles in a gauge theory, or any particles in a
theory with gravity. The continuation thus has arbitrary features
that are likely to be unpleasant.

One could imitate this approach in the context of covariant or light cone string field theory (and probably one does not
need here the full machinery of string field theory). In general,  the off-shell continuation involved might be quite unpleasant,
and it might be difficult to understand the 
analytic continuation to the physical region and to restore gauge-invariance if it has been lost in the continuation.  
Still,  something along these lines has actually been done for some 1-loop amplitudes \cite{amano}.

\subsection{The Feynman $i\varepsilon$}\label{feynman}

However, in field theory, the most direct and transparent approach to the singularity of the propagator is that of Feynman. One works directly
with on-shell momenta in Lorentz signature.  In signature $-++\dots +$ (which we choose over $+--\dots -$, as it is more directly related to the Euclidean
case),    the propagator is
\begin{equation}\label{dooz}\frac{-i}{p^2+m^2-i\varepsilon},  \end{equation}
where $\varepsilon$ is an infinitesimal positive quantity, taken to 0 at the end of a calculation.
This displaces the zero of the denominator away from the locus of real $p$, over which one integrates, and gives the most
direct way to calculate perturbative $S$-matrix elements.

The Euclidean signature propagator can be naturally written as an integral over Euclidean signature proper time  $\tt$
\begin{equation}\label{realt}\frac{1}{p^2+m^2}=\int_0^\infty \d\tt \,\exp(-\tt(p^2+m^2)). \end{equation}
This integral converges for $p^2+m^2>0$; in other words, it converges if the energy is below threshold.
The Lorentz signature Feynman propagator can similarly be written as an integral over Lorentz signature proper time $\tau$:
\begin{equation}\label{lort}\frac{-i}{p^2+m^2-i\varepsilon }=     \int_0^\infty \d\tau\,\exp(-i\tau(p^2+m^2-i\varepsilon)).       \end{equation}
Here we include a convergence factor $\exp(-\varepsilon\tau)$ in what would otherwise be an oscillatory integral.  With this convergence factor,
the integral converges for all real $p^2$, above or below the pole of the propagator.  The limit $\varepsilon\to 0$ exists pointwise except precisely at the pole,
and in general exists as a distribution.   We will refer to the Lorentz signature proper time $\tau$ in (\ref{lort})
and the Euclidean signature proper time $\tt$ in (\ref{realt}) as real and imaginary time (or Lorentz signature and Euclidean signature) Schwinger parameters, respectively.

When we represent a string theory scattering amplitude as an integral over the moduli of a Riemann surface, these modular parameters are the generalizations
of imaginary time Schwinger parameters.  For instance, in the Veneziano amplitude
\begin{equation}\label{zorok}A_{\mathrm V}(s,t)=\int_0^1\d x\, x^{-\alpha' s -2}(1-x)^{-\alpha' t-2},\end{equation}
near $x=0$, $x$ corresponds in field theory terms to $e^{-\tt}$, where $\tt$ is an imaginary time Schwinger parameter.  The Schwinger parameters are Euclidean
because the string worldsheet is Euclidean.

If we could do string perturbation theory with Lorentz signature worldsheets, then the modular parameters would be generalizations of real time Schwinger
parameters, and the Feynman $i\varepsilon$ would be natural.   However, Lorentz signature worldsheets describing string interactions would have singularities
associated to breaking and joining of strings, and we would lose the natural understanding of the absence of ultraviolet divergences that we get from the
Euclidean framework. (The only commonly considered framework in which the worldsheet can be considered to be Lorentzian is light cone gauge
\cite{Mandeltwo}.)

\begin{figure}
 \begin{center}
   \includegraphics[width=3in]{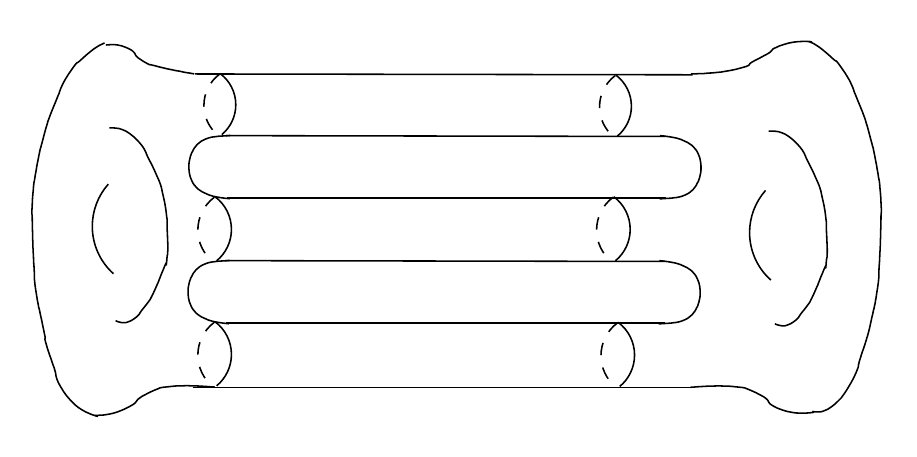}
 \end{center}
\caption{\small A string world sheet in closed-string field theory, with three propagators -- represented by simple tubes -- joining trivalent vertices that in general
are represented by rather complicated worldsheets.  (The vertices here are depicted as smooth genus 1 worldsheets, though  in closed-string field theory the vertices carry Kahler metrics
that often are not smooth.)}
 \label{built}
\end{figure}
In string field theory -- either covariant or light cone -- there is no problem with the $i\varepsilon$.   The case of covariant closed string field theory as
reviewed in \cite{Zwiebach} is instructive for our purposes.  A string worldsheet is built (fig. \ref{built}) by gluing together vertices -- which can be very complicated but
are described by Euclidean worldsheets -- with tubes that correspond to propagators.  The propagator\footnote{We omit some factors $b_0\t b_0\delta_{L_0-\t L_0}$ that
are inessential for our present purposes.} is $-i/L_0$, and by hand we can replace this with $-i/(L_0-i\varepsilon)$.  We can think of this as arising from an integral
over a real time Schwinger parameter:
\begin{equation}\label{zelb}\frac{-i}{L_0-i\varepsilon}=\int_0^\infty \d\tau \,\exp(-i\tau(L_0-i\varepsilon)). \end{equation}
So the replacement $L_0\to L_0-i\varepsilon$ amounts to saying that the tubes that represent the propagators have Lorentz signature, even though
the vertices are represented by Euclidean worldsheets.

\begin{figure}
 \begin{center}
   \includegraphics[width=2.6in]{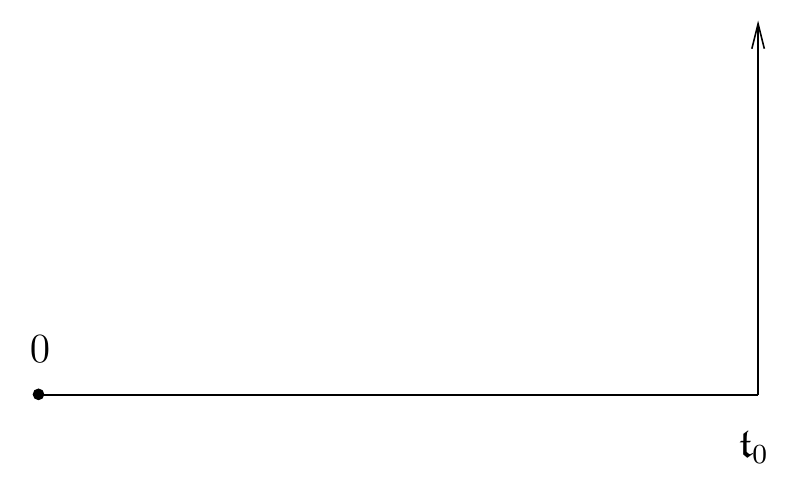}
 \end{center}
\caption{\small This contour represents integration over Euclidean proper time $\tt$ up to $\tt=\tt_0$ for some large $\tt_0$, and then continuing
with $\tt=\tt_0+i\tau$, $0\leq\tau<\infty$. }
 \label{contour}
\end{figure}

A field theory version of this would be to integrate the proper time over the sort of contour sketched in fig. \ref{contour}.  We start with Euclidean signature proper time $\tt$
and integrate on the real axis.  But when $\tt$ becomes large -- which is where the integral may fail to converge -- we make a 90 degree turn.  We integrate over
the real $\tt$ axis up to $\tt=\tt_0$ (for some $\tt_0$ that may be large), and beyond this we set $\tt=\tt_0+i\tau$
where now $\tau$ is a Lorentz signature proper time parameter that varies from 0 to $\infty$.  With the help of a convergence factor $\exp(-\varepsilon\tau)$,
the integral on this contour converges for all real $p^2$, and gives the Feynman propagator with the $i\varepsilon$.

\begin{figure}
 \begin{center}
   \includegraphics[width=3.3in]{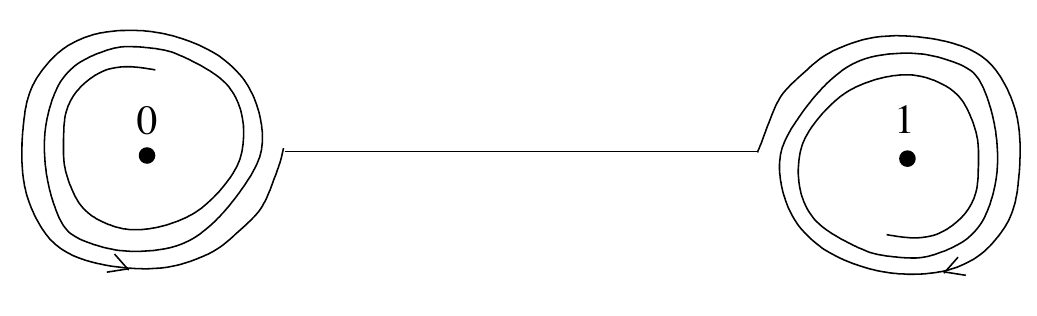}
 \end{center}
\caption{\small An integration contour for the Veneziano amplitude that incorporates the Feynman $i\varepsilon$.  The part of the contour near $x=0$ and 1 is drawn
as a tight spiral to make it legible, but actually (to ensure convergence whenever $s,t,$ and $u$ are real)
one wants to repeat the same little circle around 0 or 1  infinitely many times, with the help of an $\exp(-\varepsilon\Phi)$ convergence factor. }
 \label{depicted}
\end{figure}
We actually can easily implement this for the Veneziano amplitude.  As in section \ref{pochhammer}, we start with the fact that the form $\omega=\d x\, x^{-\alpha' s-2}
(1-x)^{-\alpha't-2}$ that we wish to integrate has an analytic continuation to the complex $x$-plane, with branch points (of infinite order) at 0 and 1.  We want to find an
integration contour for the $x$ integral that matches what we said in the field theory case.  We just need to slightly modify the standard contour near 0 and 1 to get an
integral whose convergence is not a problem.  Near $x=0$, we think of $x$ as $e^{-\tt}$, where when $\tt$ reaches a large value $\tt_0$, we want to continue with $\tt
=\tt_0+i\tau$, which means that $x$ will be $e^{-\tt_0}e^{-i\tau}$.  So we integrate over real $x$ down to a very small value and thereafter we revolve forever, in a clockwise
direction, around $x=0$.  We do the same thing near $x=1$, and thus the integration contour is as depicted in fig. \ref{depicted}.   

Just as in Feynman's case, we need to supply a convergence factor. 
All that is important here -- as in field theory -- is that the convergence factor is smooth,  vanishes at the ``ends'' of the integration contour for 
any $\varepsilon>0$, and approaches 1 pointwise for $\varepsilon\to 0$.  With the help of this integration factor, the integral defining the Veneziano amplitude converges
for all real $s,t,u$.  In this paper, we generically write such convergence factors as $\exp(-\varepsilon \Phi)$.  Though not strictly necessary, it is convenient, in order to match with field
theory, to assume
that at infinity along the integration contour, 
$\Phi$ grows linearly with the proper time.
 For instance, in the present example, we can take the convergence factor to
be  $\exp(-i\varepsilon(\log x+\log (1-x)))$.    In our later examples, the integration cycle is multidimensional and we take $\Phi$ to grow linearly with the sum of the proper times.

We can do something similar for closed string amplitudes.  For example, in the case of the Virasoro-Shapiro amplitude, near $z=0$, we introduce polar coordinates
$z=\rho e^{i\varphi}$, $\t z=\rho e^{-i\varphi}$.  The usual integration cycle is defined by $\t z =\bar z$ or equivalently by real $\rho,\,\varphi$.  To incorporate the Feynman
$i\varepsilon$, we leave $\varphi$ real and but treat $\rho$ precisely as we treated $x$ in the case of the Veneziano amplitude.  This means that we integrate
on the real $\rho$ axis down to $\rho=\exp(-\tt_0)$ for some large $\tt_0$, and thereafter continue with $\rho=\exp(-\tt_0-i\tau)$, with $0\leq \tau<\infty$.  
The convergence factor is now something like $\exp(-\varepsilon\tau)$ (or $\exp(-i\varepsilon \log(z\t z)/2))$.  We follow a similar procedure near $z=1$ and $z=\infty$.

In each of these examples, whenever the original integral converged at a given endpoint or singular point ($x=0,1$ or $z=0,1,\infty$), the more elaborate procedure
that we have described does not affect the contribution near that endpoint. Thus
this procedure  agrees with the original definitions of the Veneziano and Virasoro-Shapiro amplitudes to the extent that those make sense.  It is physically
sensible because it has been chosen to produce the correct poles associated to any on-shell state, properly shifted by $i\varepsilon$.  

Compared to the Pochhammer contours described in section \ref{pochhammer}, this procedure gives a satisfactory definition of the amplitudes
in question  only when the kinematic invariants are all real (for instance, when the external momenta are real in Lorentz signature).  That is a drawback, but in return we gain two advantages: the procedure
described here incorporates the Feynman $i\varepsilon$ and -- as we discuss next -- it
 generalizes straightforwardly to  loop amplitudes.  It is natural for these two virtues to go together, since in field theory 
 the Feynman $i\varepsilon$ is needed to make sense of loop
amplitudes.  

\subsection{Generalization}\label{genamp}

There is actually no difficulty in generalizing what we have said about tree-level 4-point functions in section \ref{feynman}  to an arbitrary
perturbative string theory amplitude. 

\begin{figure}
 \begin{center}
   \includegraphics[width=5.5in]{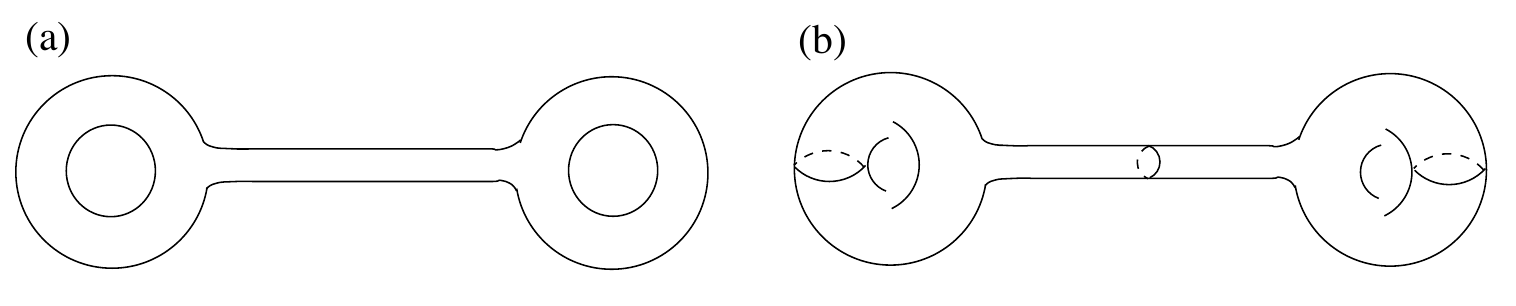}
 \end{center}
\caption{\small ``Infinity'' in string perturbation theory is the region where the proper time variable of an open string (a) or a closed string (b) becomes large.  In this region, the worldsheet
separates at least locally into two parts connected by a long strip or long tube. }
 \label{assoc}
\end{figure}
This is actually true for the same reason that there are no ultraviolet divergences in string perturbation theory.  
The moduli space of Riemann surfaces is compact except
for the ``long strip'' and ``long tube'' regions that are associated to on-shell open and closed strings, respectively
(fig. \ref{assoc}).  We have already encountered such regions in studying the Veneziano and Virasoro-Shapiro amplitudes, and they can always be treated as we have done in those cases.
Whenever a string worldsheet develops a long strip or long tube
-- so that we run into what in field theory would be the pole of the propagator --
we treat the proper time as Euclidean until it is very large, whereupon
we continue to Lorentzian proper time.  When several long strips or long tubes appear at once, we do the same thing for each one independently.
  This procedure suffices to incorporate the Feynman $i\varepsilon$ in arbitrary string theory amplitudes.

Let us spell this out in a little more detail.
An almost on-shell open string has a Euclidean Schwinger parameter $\tt$.  As we did for the tree-level four-point function, we integrate
over real $\tt$ up to a cutoff $\tt_0$, and beyond that set $\tt=\tt_0+i\tau$ with $0\leq \tau<\infty$.  In addition to a Schwinger parameter $\tt$, an almost on-shell closed
string has a twist angle $\varphi$. ($\tt$ and $\varphi$ are the length and twist parameters of the long tube that describes the propagation of the closed string.)
 We leave the integral over $\varphi$ untouched and treat the integral over $\tt$ exactly as was just described for open strings.  
 In each case,
we equip the integral over $\tau$ with a convergence factor, such as the familiar $\exp(-\varepsilon\tau)$.    For a detailed example, see section \ref{treefive}
below or ref. \cite{berera}.

\subsubsection{Where Does The Integration Cycle Live?}\label{where}

By means of this analytic
continuation, we find what we claim is the right cycle on which we should integrate to define a perturbative string theory amplitude.  But in what space is this
cycle defined?  Clearly, in some sense it lives in the complexification of the naive moduli space.  We will now explain this more fully.

First we consider bosonic open and/or unoriented strings.  (See, for example, section 7 of \cite{wittensurfaces} for background. For the generalization of
what follows to superstring theory, see section \ref{another} below.) An open and/or unoriented string worldsheet $\Sigma$ has a closed oriented double cover
$\hat\Sigma$ that is a Riemann surface of genus $g$ with $n$ punctures, for some $g$ and $n$.  Let $\M_{g,n}$ denote the moduli space of such surfaces.  $\M_{g,n}$ parametrizes the possible deformations of $\hat\Sigma$.  It has an involution (in other
words a $\Z_2$ symmetry) of ``complex 
conjugation,'' which we will call $\bm\tau$.  ($\bm\tau$ maps a Riemann surface with a given complex structure to the same surface with the opposite complex
structure.) The fixed point set of $\bm\tau$ has in general many components. One component of this fixed point set is the
moduli space $\Gamma$ that parametrizes the possible variations of the original open and/or unoriented surface $\Sigma$.  (Other components
correspond to other topological types of open and/or unoriented worldsheet with the same Euler characteristic as $\Sigma$.)

The facts just stated imply that we can regard $\M_{g,n}$ as a natural complexification of the moduli space $\Gamma$ on which we naively would integrate
to compute the contribution of the open and/or unoriented surface $\Sigma$ to a scattering amplitude.  Hence, one's first thought may be that the
corrected integration cycle that incorporates the Feynman $i\varepsilon$ would lie in $\M_{g,n}$.  This is not quite the right answer for a reason that we saw
in our discussion of the Veneziano amplitude.  The form on $\M_{g,n}$ that one gets by analytic continuation of the open and/or unoriented string theory
path integral from $\Gamma$ to $\M_{g,n}$ is multi-valued.  To define a sensible integration cycle, one must replace $\M_{g,n}$ by a cover on which
the form that one wishes to integrate is single-valued.   (This cover still has $\Gamma$ as the fixed point set of an involution, so it 
can still be viewed as a complexification of $\Gamma$, though not a minimal one.)

In a general string theory compactification, to make the integration form single-valued, 
it may be necessary to pass from $\M_{g,n}$ to its universal cover, which is the Teichmuller space $\T_{g,n}$.
However, in the specific case of bosonic strings in $\R^{26}$, one can make do with a smaller cover of $\M_{g,n}$, namely its maximal abelian cover
(in which the covering group is\footnote{In general, $\M_{g,n}$ must be regarded as an orbifold and its homology and fundamental group must be defined in the orbifold sense.} $H_1(\M_{g,n},\Z)$, as opposed to $\pi_1(\M_{g,n})$, which would be the covering group of the universal cover).  Let us see how this works for the 
tree-level $n$ tachyon amplitude.  As in eqn. (\ref{gurf}), this amplitude is formally defined by
\begin{equation}\label{gurfy}A_{n,\mathrm{open}}=\frac{1}{\vol\, SL_2(\R)}\int_S \d x_1\dots \d x_n\prod_{i<j} |x_i-x_j|^{2\alpha' p_i\cdot p_j} .\end{equation}
To incorporate the Feynman $i\varepsilon$, we first analytically continue from real to complex values of the $x_i$.  In the process, the information
about the cyclic ordering of the $n$ points is lost, and the group $SL_2(\R)$ is replaced by $SL_2(\C)$.  The resulting moduli space is $\M_{0,n}$,
the moduli space of $n$ points, parametrized by the complex variables $x_i$, in $\CP^1$.  The form that we want to integrate to define the amplitude
is now 
\begin{equation}\label{zob}\omega=\frac{1}{\vol\, SL_2(\C)}\d x_1\dots \d x_n \prod_{i<j}(x_j-x_i)^{2\alpha' p_i\cdot p_j}. \end{equation}
This form is multi-valued, with branch points whenever $x_i=x_j$ for some $i,j$.  However, since the monodromies around the branch points
are abelian (under monodromy, the form $\omega$ is simply multiplied by a complex constant), 
to find a space in which $\omega$ is single-valued, it suffices to pass to the
maximal abelian cover $\t\M_{0,n}$ of $\M_{0,n}$.  The integration cycle $\t \Gamma$ of the tree-level $n$ tachyon amplitude can therefore be taken
to lie in $\t\M_{0,n}$.  

For example, for $n=4$ -- that is, for the Veneziano amplitude -- after fixing the $SL_2(\R)$ symmetry, the form that should be single-valued on $\t\M_{0,4}$ is
our friend $\omega=\d x \,x^{-\alpha' s-2}(1-x)^{-\alpha' t -2}$.  To find a cover of the complex $x$-plane on which this form is single-valued, we simply
introduce new variables $u$ and $v$ with $e^u=x$, $e^v=1-x$, so
\begin{equation}\label{zorg} e^u+e^v=1.  \end{equation}
(Complex curves described by equations of this type are important in mirror symmetry \cite{vh}; it is not clear if this is a coincidence.)
The integration contour $\t\Gamma$ runs from $u=-u_0-i\infty$ (with $u_0$ a large constant and $v$ near 1) to $v=-v_0-i\infty$ (with $v_0$ a large
constant and $u$ near 1).  The analog of this for $n>4$ can be described similarly, though more variables and equations are required.

Now let us discuss the analog of this in closed oriented string theory.  Consider the contribution in string perturbation theory of
 a closed oriented Riemann surface $\Sigma$ of genus $g$
with $n$ punctures, for some $g$ and $n$.  Its deformations are parametrized by the corresponding moduli space $\M_{g,n}$.  $\M_{g,n}$ is
of course naturally a complex manifold of dimension $3g-3+n$.  However, to incorporate the Feynman $i\varepsilon$, we want to view $\M_{g,n}$
as a real manifold of dimension $6g-6+2n$ and complexify it so that local holomorphic coordinates on $\M_{g,n}$ and their complex conjugates
become independent complex variables.  The natural complexification of $\M_{g,n}$  is simply a product $\X=\M_L\times \M_R$ where $\M_L$
and $\M_R$ are two copies of $\M_{g,n}$, but with reversed complex structure on $\M_L$.  (The motivation for the names $\M_L$ and $\M_R$
is that $\M_L$ and $\M_R$ parametrize, respectively, the complex structures that are ``seen'' by left- and right-moving or antiholomorphic and holomorphic
degrees of freedom on $\Sigma$.)  $\X=\M_L\times \M_R$ has an antiholomorphic involution $\bm\tau$ that exchanges the two factors (this exchange
is antiholomorphic since we have reversed the complex structure on $\M_L$), and the fixed point set of $\bm\tau$ is the diagonal, which is a copy of 
$\M_{g,n}$.    To state this in reverse, $\X$ is a complexification of $\M_{g,n}$.  We can take the integration cycle that incorporates the Feynman $i\varepsilon$
to lie in a cover of $\X$ on which the form that has to be integrated is single-valued.  (This cover can still be viewed as a complexification of $\M_{g,n}$.)

We can make all this explicit for the tree-level amplitude of $n$ closed-string tachyons.    As in eqn. (\ref{wurf}), this amplitude is naively defined by an
integral over $\M_{0,n}$:
\begin{equation}\label{wurfy}A_{n,\mathrm{closed}}=\frac{1}{\vol \,SL_2(\C)}\int \d^2z_1\dots \d ^2z_n\prod_{i<j} |z_i-z_j|^{\alpha' p_i\cdot p_j} .\end{equation}
To find the space on which the integration cycle is defined, first we complexify $z_i$ and its complex conjugate $\bar z_i$ to independent complex variables $z_i$ and $\t z_i$; we also complexify\footnote{As we explained above with the example of $\M_{g,n}$, if $U$ is a complex manifold then $U\times U$, with the opposite complex
structure on the first factor, can be viewed as a complexification
of $U$.}
$SL_2(\C)$ to $SL_2(\C)\times SL_2(\C)$, with one copy acting on the $z_i$ and one on the $\t z_i$.  Thus the integral (\ref{wurfy}) runs over
 a cycle $\Gamma\subset \M_{0,n}\times \M_{0,n}$ defined by $\t z_i=\bar z_i$; $\Gamma$ is simply a diagonal 
 copy of $\M_{0,n}\subset\M_{0,n}\times \M_{0,n}$. To incorporate the Feynman $i\varepsilon$, we must modify
 the integration cycle near the points $z_i=z_j$ so that $\t z_i$ is no longer the complex conjugate of $z_i$.  To get the right modification, 
 we must replace $\M_{0,n}\times\M_{0,n}$
 by a cover on which the form that must be integrated, namely
 \begin{equation}\label{donk}\frac{1}{\vol\,SL_2(\C)} \d z_1\dots \d z_n \prod_{i<j}(z_i-z_j)^{\frac{\alpha'}{2}p_i\cdot p_j}\cdot 
 \frac{1}{\vol\,SL_2(\C)} \d \t z_1\dots \d \t z_n \prod_{i<j}(\t z_i-\t z_j)^{\frac{\alpha'}{2}p_i\cdot p_j},\end{equation}
 is single-valued. We can take this to be an abelian cover, since the monodromies of the form in question are abelian.  The integration cycle $\t \Gamma$
 that incorporates the Feynman $i\varepsilon$ lives in this abelian cover.   The cover can be described explicitly in terms of variables analogous to $u$ and $v$
 in (\ref{zorg}).

\subsubsection{Another Reason For Complexification}\label{another}

The reader may be surprised that to define perturbative string amplitudes properly, one must integrate not over what naively is the moduli space of 
string worldsheets but over a suitable cycle in an appropriate complexification of this space.  However, there is another reason that this is necessary.

So far in this paper, we have not mentioned worldsheet or spacetime supersymmetry; they are not relevant in a general discussion of the Feynman $i\varepsilon$.
However, a perhaps little-appreciated difference between bosonic string theory and superstring theory is that, apparently, there does not
exist a natural moduli space of superstring worldsheets.  This has been explained in section 5 of \cite{wittenintegrate} and illustrated in elementary terms in
section 2.4.4 of \cite{more}.  Although it seems that a natural moduli space of superstring worldsheets does not exist, there is no problem in constructing
a suitable stand-in for the complexification of this space.  For closed oriented superstrings, this stand-in 
is simply a product $\X=\M_L\times \M_R$, where $\M_L$ and $\M_R$ parametrize the moduli ``seen''
by antiholomorphic and holomorphic degrees of freedom on the string worldsheet.  For example, for the heterotic string, $\M_R$ is a moduli space of super
Riemann surfaces and $\M_L$ is a corresponding moduli space of ordinary Riemann surfaces; for Type II superstrings, $\M_R$ and $\M_L$ are both
moduli spaces of super Riemann surfaces.\footnote{The analog for open and/or unoriented superstrings is as follows (see section 7 of \cite{wittensurfaces}).  
The stand-in for the complexification of the would-be moduli
space of open and/or unoriented superstring worldsheets is a copy of the moduli space of super Riemann surfaces, parametrizing the deformations of the
closed oriented double cover of the worldsheet.}  $\X$ is a stand-in for the complexification of a hypothetical moduli space of superstring worldsheets since
if such a moduli space did exist, one would expect it to have $\X$ as a complexification, by analogy with what we have explained for  bosonic strings.

Superstring perturbation theory is based on integration over
an appropriate cycle $\varGamma\subset \X$.  As explained in the above-cited references, there is no natural choice of $\varGamma$, but it is naturally
determined up to homology, in fact up to nilpotent deformations.  Because the form that must be integrated to compute a superstring scattering amplitude is
closed, a knowledge of $\varGamma$ up to homology would suffice to determine an amplitude if $\varGamma$ were compact.  Since $\M_L$, $\M_R$,
and $\varGamma$ are all noncompact, one actually needs to supply some information about the behavior of $\varGamma$ at infinity in order to 
make superstring perturbation theory well-defined.

\begin{figure}
 \begin{center}
   \includegraphics[width=5in]{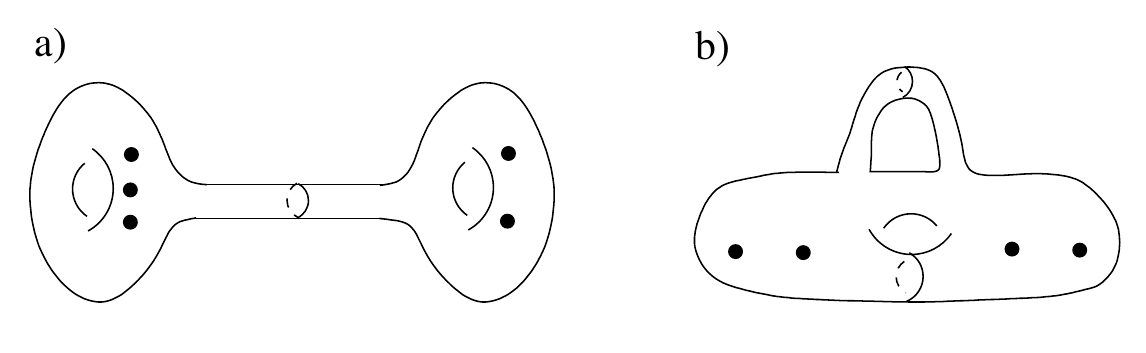}
 \end{center}
\caption{\small Generic degenerations in which the momentum flowing through a degenerating cycle, represented by a long narrow tube,
 is generically off-shell.  Depicted in (a) is  a separating degeneration.  The degenerating
cycle connects the left and right parts of the worldsheet; the momentum flowing through it is determined by the external momenta.  Depicted in (b) is a nonseparating degeneration (cutting the long
narrow tube does not separate the Riemann surface into disjoint pieces);
from a field theory point of view, the momentum flowing through the long tube  is an arbitrary loop momentum variable.  }
 \label{offshell}
\end{figure}
\begin{figure}

 \begin{center}
   \includegraphics[width=3in]{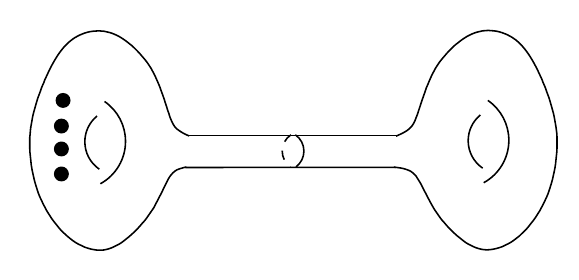}
 \end{center}
\caption{\small A special degeneration in which the momentum flowing through the degenerating cycle -- which separates the left and right portions of the worldsheet -- is automatically on-shell for some
string state.
In the example given here, this is true because the external particles are all on the left side, and momentum conservation implies that their momenta add to 0.  So a massless string state propagating
down the long tube is automatically on-shell.}
 \label{on-shell}
\end{figure}
There are two distinct cases of this problem.  ``Infinity'' in moduli space always refers to the limit in which a cycle on the string worldsheet degenerates
(or alternatively, in a different conformal frame, the limit that an open or closed string propagates for a long proper time).  
However, the two cases in which the momentum flowing through the degenerating cycle
is generically not on-shell (fig. \ref{offshell}) or is generically on-shell (fig. \ref{on-shell}) are quite different.      

When the momentum flowing through a degeneration
is generically not on-shell, we say that this is a generic degeneration.  In this case,
the Feynman $i\varepsilon$ comes into play: it controls the singularity that arises when the momentum flowing through the
degenerating cycle goes on-shell.  
(That is so whether this momentum is determined by the external momenta as in fig. \ref{offshell}(a) or from a field theory point of view depends on loop momenta
as in fig. \ref{offshell}(b).)
The present paper has been devoted to generic degenerations.  Worldsheet and spacetime supersymmetry have played no particular
role in analyzing them.  Nor, in the case of superstring theory, does the GSO projection play any particular role.
The Feynman $i\varepsilon$ applies alike to GSO-even and GSO-odd states, although the contributions of the GSO-odd states to physical singularities
are removed when one performs the complete integral over $\varGamma$ (this integral includes near each degeneration a partial sum over spin structures
that implements the GSO projection; see for instance section 6.2.3 of \cite{wittensurfaces}).

When the momentum flowing through a degeneration is generically on-shell, so that $p^2+m^2=0$, we say that this is a special degeneration.
In this case,
the propagator $-i/(p^2+m^2-i\varepsilon)$ is simply $1/\varepsilon$ and the Feynman $i\varepsilon$ does not help very much; there is no limit for $\varepsilon\to 0$ unless
something cancels the pole of the propagator. 
These special degenerations cause bosonic string 
perturbation theory to break down at the 1-loop level and lead 
to the greatest subtlety in superstring perturbation theory. Spacetime supersymmetry is definitely important in taming them.  
Dealing with those special degenerations was the main goal in a recent reconsideration of superstring perturbation theory \cite{revisited}.
In analyzing them, a cutoff procedure was used in which worldsheet supersymmetry played an important role.
The procedure was rather different from the one we have followed in the present paper for  generic degenerations.
It is not clear to the author if the two types of degeneration could be treated more uniformly.

One moral of this story is that there are two somewhat independent reasons that string perturbation theory really has to be formulated in terms of integration
not over a moduli space of string worldsheets but over a suitable cycle in the complexification of such a space.  
One reason is that in the case of superstring theory, only
the complexification and not the naive moduli space exists, as explained in \cite{wittenintegrate,more}.  The second reason is that, as described
in the present paper, even in bosonic string theory, replacing the naive space with a suitable cycle in its complexification
is needed to incorporate the Feynman $i\varepsilon$.   

\subsubsection{More On The Schwinger Parameters And Their Extension To Superstring Theory}\label{superext}
 
In general, an open-string degeneration (fig. \ref{assoc}(a)) of bosonic strings corresponds to a boundary of the moduli space $\Gamma$ of open-string worldsheets.
This degeneration can be described\footnote{The degeneration is described by an equation $xy=q$, where $x$ and $y$ are local
parameters on two Riemann surfaces (or two branches of the same Riemann surface) that can be glued together at $q=0$ to make a singular Riemann
surface, or deformed to a smooth one with $q\not=0$.  For details on this and on other matters that will be mentioned momentarily,
 see for example sections 6 and  7 of \cite{wittensurfaces}.} by a gluing parameter $q$.  $q$ is a non-negative function on $\Gamma$ that has a simple zero along the boundary.
As such it is defined modulo
\begin{equation}\label{zelbo} q\to e^f q ,\end{equation} 
where $f$ is any function that is regular along the boundary of $\Gamma$, that is at $q=0$.
We define the Euclidean Schwinger parameter $\tt$ for this degeneration by $q=e^{-\tt}$.  In view of (\ref{zelbo}), this means that $\tt$ is defined
modulo the possibility of adding to it a function that is bounded for $\tt\to\infty$.  Adding to $\tt$ such a function does not affect any of our considerations
in any way.

A closed-string degeneration (fig. \ref{assoc}(b))  corresponds to a complex divisor in the moduli  space of Riemann surfaces.  It is defined locally by
the vanishing of a  holomorphic function $q$, which again can be interpreted as a gluing parameter, and which again
is subject to an indeterminacy analogous to (\ref{zelbo}). We define the Schwinger parameter
$\tt$ by setting $q=e^{-\tt+i\phi}$, with real $\tt$, $\phi$.  Again, $\tt$ is uniquely defined modulo the addition of a function that is bounded for $\tt\to\infty$.

All this has a close analog in superstring theory.  Open- and closed-superstring degenerations are controlled by  parameters analogous to the bosonic 
gluing parameters mentioned in the last two paragraphs.  An open-superstring degeneration is parametrized locally by an even function $q$ that is real and positive
modulo the odd variables, and a closed-superstring degeneration is parametrized locally by an even holomorphic function $q$.  
These parameters are again gluing parameters, though a proper explanation involves some subtleties of super Riemann surfaces (see for example sections 6 and 7 of
\cite{wittensurfaces}).
The degenerations occur again for $q\to 0$, and the $q$'s again have an indeterminacy
 $q\to e^fq$, where $f$ is regular at $q=0$.  
One again defines a Schwinger parameter  in terms of $q$ by $q=e^{-\tt}$ for open strings or $q=e^{i\varphi}e^{-\tt}$ for closed strings.   
The indeterminacy $q\to e^f q$ again means that $\tt$ is defined modulo addition of a function that is bounded at infinity, but now we need to clarify the meaning of this
statement.  If for example $\eta_1$ and $\eta_2$ are odd moduli of the limiting surface at $q=0$, then we say that $\tt +a(\tt)+b(\tt)\eta_1\eta_2$ differs from $\tt$ by a bounded
function if and only if  the functions $a(\tt)$ and $b(\tt)$ are both bounded for $\tt\to\infty$.

At a Ramond degeneration, in addition to $q$, there is also a fermionic gluing
parameter which plays an important role in turning the propagator from $1/L_0$ to $G_0/L_0$ (where $G_0$ is a global supercharge).  But it does not affect
the definition of the Schwinger parameter and so does not affect our discussion here.

 Importantly, the indeterminacy $q\to e^fq$  
means that if $\eta_1$ and $\eta_2$ are odd variables (which technically are moduli of the limiting surface at $q=0$), 
we are free to transform $q\to q(1+\eta_1\eta_2)$, but not $q\to q+\eta_1\eta_2$.
In the case of
the special degenerations of fig. \ref{on-shell}, if one uses $q+\eta_1\eta_2$ as a variable instead of $q$, one will in general get  wrong answers;
for instance, see section 2 of \cite{more}.  The present paper is really concerned with generic degenerations.
At a generic degeneration, using the right gluing parameters
in the definitions of the proper time variable $\tt$ is important in matching naturally with field theory.  See section 6 of \cite{revisited} for an
explanation of how the usual Feynman propagators of field theory emerge from string theory if one computes using the right gluing parameters.
See also \cite{sciuto} for an analysis of a generic two-loop open-string degeneration with a detailed match to field theory.

It is instructive to see explicitly why it is important that we are allowed to replace the gluing parameter $q$ by $q(1+\eta_1\eta_2)$ and not by $q+\eta_1\eta_2$.
Set $q=e^{-\tt}$, $q'=e^{-\tt'}$.  If $q'=q(1+\eta_1\eta_2)$, then $\tt'=\tt-\eta_1\eta_2$, so the difference between $\tt$ and $\tt'$  is bounded at infinity and it does
not matter which one we use as a proper time variable in comparing to field theory or introducing the $i\varepsilon$ prescription.  But if $q'=q+\eta_1\eta_2$, then $\tt'=\tt-\eta_1\eta_2 e^\tt$, and the difference between $\tt$ and $\tt'$
is definitely not bounded.  If we get a good match to field theory with $\tt$ as the Schwinger parameter, this will not work if we try to use $\tt'$.

\subsection{Five-Point Function At Tree Level}\label{treefive}

Here we will describe in somewhat more detail  the integration cycle that incorporates the Feynman $i\varepsilon$ 
for the case of a 5-point tree amplitude of bosonic open strings.  This will illustrate when happens when 2 or more degenerations can occur simultaneously.
Once one understands what the integration cycle looks like in this case, there are no surprises in higher examples, including loop amplitudes.
(The reader may want to contrast what we will describe with the Pochhammer-like cycle described for the same amplitudes in \cite{hsh}.)

\begin{figure}
 \begin{center}
   \includegraphics[width=2in]{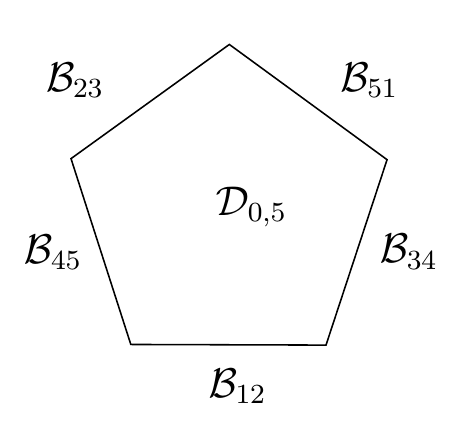}
 \end{center}
\caption{\small The moduli space $\D_{0,5}$ can be depicted as a pentagon.  Its five boundary components correspond to the five different channels in which a proper time
parameter may go to infinity, and the corners correspond to the five ways that a pair of proper time parameters can go to infinity.}
 \label{pentagon}
\end{figure}

The 5-point tree amplitude is formally
an integral over  the moduli
space $\D_{0,5}$ that parametrizes 5 cyclically ordered points $x_1,\dots,x_5$  on the boundary of a disc, modulo the action of $SL_2(\R)$. (Because the points are only cyclically ordered, we consider
the subscript $i$ of $x_i$ to be defined mod 5.)  $\D_{0,5}$ has real dimension 2.  It has 5 boundary components\footnote{These boundary components  lie
really not in $\D_{0,5}$ but in a natural
compactification of this space, an open-string analog of the Deligne-Mumford compactification of the moduli space of Riemann surfaces.} $\B_{i,i+1}$, $i=1,\dots, 5$.
$\B_{i,i+1}$ parametrizes limiting configurations that arise 
for $x_i$ and $x_{i+1}$ becoming coincident.   The moduli space 
$\D_{0,5}$ can thus be pictured as a pentagon (fig. \ref{pentagon}),
a two-dimensional convex polygon with 5 boundary components.  There is no additional boundary component with, say, $x_1,x_2,x_3$ all becoming
coincident, because up to an $SL(2,\R)$ transformation, this is equivalent to $x_4\to x_5$, which corresponds to one of the 5 boundary components that we have
already identified.

On the other hand, pairs of boundary components of $\D_{0,5}$ can intersect, since for example it is possible to have $x_1\to x_2$ simultaneously
with $x_4\to x_5$.  (This can be done while fixing the $SL_2(\R)$ symmetry by keeping fixed $x_2,x_3,x_4$.)  So $\B_{12}$ intersects $\B_{45}$ and
similarly intersects $\B_{34}$.  The intersections of boundary components
are represented by the corners of the pentagon in fig. \ref{pentagon}.  

\begin{figure}
 \begin{center}
   \includegraphics[width=5in]{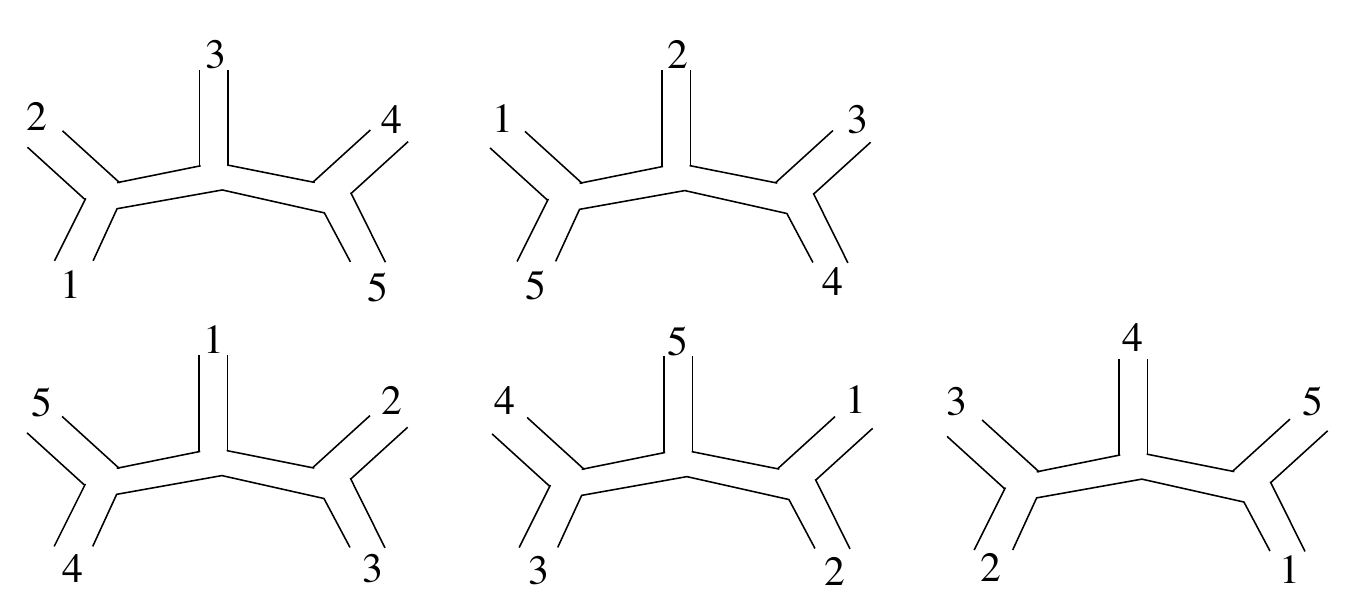}
 \end{center}
\caption{\small In cubic open-string field theory, the 5-particle tree graph (with a given cyclic ordering of the external particles) is computed by summing these 5 diagrams.}
 \label{fivetri}
\end{figure}
\begin{figure}
 \begin{center}
   \includegraphics[width=2.2in]{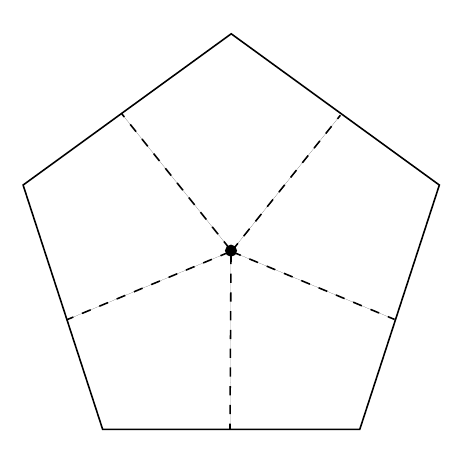}
 \end{center}
\caption{\small The 5 graphs of fig. \ref{fivetri} combine to give a covering of $\D_{0,5}$. Each graph contributes one of the 5 pieces indicated here.  This picture shows, for instance,
that each boundary component $\B_{i,i+1}$ receives contributions from two different graphs.  The reason for this is shown in fig. \ref{bycle} below. }
 \label{triangulation}
\end{figure}
\begin{figure}
 \begin{center}
   \includegraphics[width=4.5in]{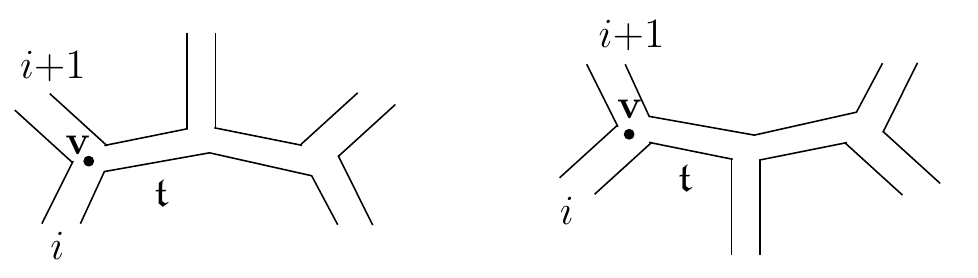}
 \end{center}
\caption{\small   The two graphs in which external strings $i$ and $i+1$ are attached to the same trivalent vertex, which we label $\mathbf v$.  We denote as $\tt=\tt_{i,i+1}$ the proper time parameter of
the internal string that attaches to $\mathbf v$.}
 \label{bycle}
\end{figure}

We can build up the same picture from open-string field theory with cubic vertices \cite{openstring}.   For scattering of 5 cyclically ordered open strings,
there are 5 possible trivalent tree graphs  (fig \ref{fivetri}).  Each such graph has 2 internal strips representing propagating strings.
Each strip has a length parameter $\tt$, which we interpret as a Euclidean Schwinger parameter.
The 2 parameters of any one graph parametrize a portion of $\D_{0,5}$, and (as usual in cubic open-string field theory \cite{gmw})
the 5 graphs together parametrize all of $\D_{0,5}$. This is shown in fig. \ref{triangulation}. To see a boundary component $\B_{i,i+1}$ in this description, we look for a graph (fig. \ref{bycle}) in which particles
$i$ and $i+1$ are attached to the same vertex $\mathbf v$. (There are 2 such graphs and each describes part of $\B_{i,i+1}$, as sketched in fig. \ref{triangulation}.) The internal strip that attaches to $\mathbf v$ has a length or proper time parameter $\tt_{i,i+1}$.
The boundary component $\B_{i,i+1}$ corresponds to the limit $\tt_{i,i+1}\to\infty$.  The 5 corners of the pentagon in fig.
\ref{pentagon}  represent intersections of boundary components, so they correspond to the points at which both proper time parameters of the same 
 graph go to infinity.

\begin{figure}
 \begin{center}
   \includegraphics[width=3in]{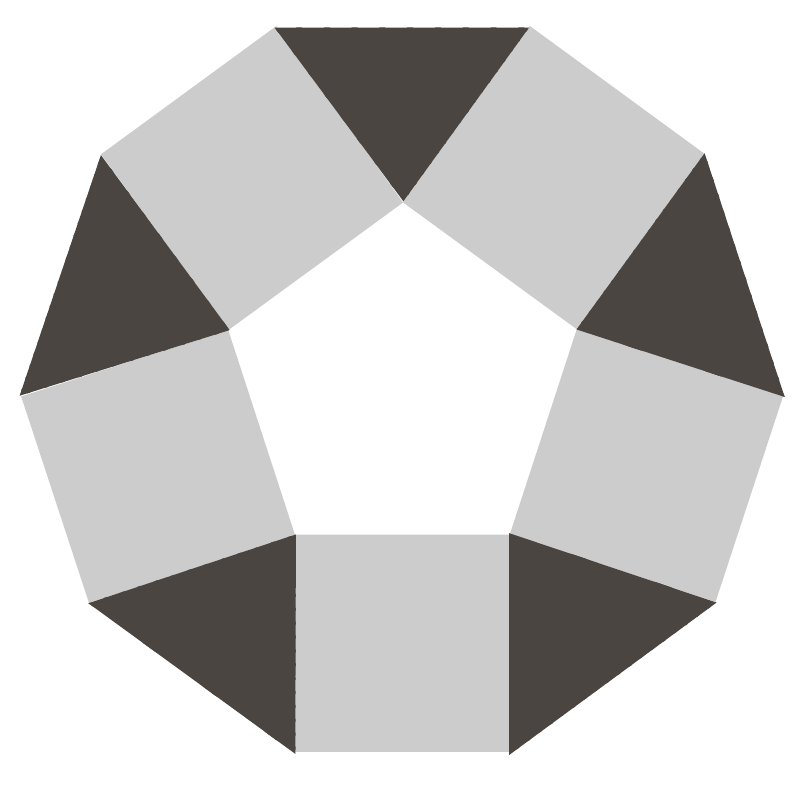}
 \end{center}
\caption{\small The integration cycle for the five open strings at tree level.  In the pentagon in the center, the Schwinger parameters are all Euclidean.
The five lightly shaded regions have 1 Lorentz signature Schwinger parameter, and the 5 more darkly shaded regions have 2 Lorentz signature Schwinger parameters.
The shaded regions go off to infinity -- though only a finite portion is drawn -- and the integration cycle is topologically a copy of $\R^2$.}
 \label{cycle}
\end{figure}

To construct the integration cycle $\tilde \Gamma$, we first pick a very large upper cutoff $\tt_0$ on all of the proper time variables arising from any graph. 
$\tilde \Gamma$ contains the portion of $\D_{0,5}$ in which
no proper time parameter  is greater than $\tt_0$.  To construct $\tilde\Gamma$, we omit from $\D_{0,5}$ the region
in which any given $\tt$ exceeds $\tt_0$, and instead we allow that variable to acquire an imaginary part in the usual way with $\tt=\tt_0+i\tau$, 
$0\leq\tau<\infty$.  Thus, for each boundary component $\B_{i,i+1}$,
we omit from $\D_{0,5}$ a small neighborhood of $\B_{i,i+1}$ and instead add a region in which the corresponding Schwinger parameter has  real
part $\tt_0$ and arbitrary positive imaginary
part.  When 2 of the $\tt$'s reach the upper cutoff $\tt_0$ (which happens near the corners of the pentagon where  boundary components intersect), we continue
each one into the complex plane along the half-line  $\tt_0+i\tau$, $\tau\geq 0$.     Accordingly, in the desired
integration cycle $\tilde\Gamma$, there are 5 regions in which one Schwinger parameter has an imaginary part and
5 more in which a pair of Schwinger parameters have imaginary parts.  The whole picture is sketched in fig. \ref{cycle}.

Importantly, as this figure shows, the different pieces of $\t\Gamma$ fit together to make a copy of $\R^2$, with no boundaries and only an end at infinity at which one or more proper
time variables go to infinity.  In its interior, $\t\Gamma$ contains a slightly truncated version of the original pentagon. In integrating over $\t\Gamma$
to compute a 5-point function, 
as long as all kinematic invariants are real, the  integral over the exterior of the pentagon is oscillatory.  
The analog of the Feynman
$i\varepsilon$ is a convergence factor $\exp(-\varepsilon\Phi)$, where $\Phi$ is any function on $\t\Gamma$ that grows at infinity (it is usually convenient to
assume that $\Phi$ grows linearly with the proper times).  If the original integral
over $\D_{0,5}$ is convergent near a given boundary component $\B_{i,i+1}$, then the part of $\t\Gamma$ in which $\tt_{i,i+1}$ has an imaginary part does
not contribute in the limit that  $\tt_0$ is large and $\varepsilon\to 0$.  

Cubic open-string field theory was a convenient guide in our reasoning but not really essential. 
 The only really essential ingredient was a qualitative understanding of the degenerations of
the classical moduli space $\D_{0,5}$.  
We have considered a tree-level process, but the same ideas apply in loops.  The main idea is always to treat each degeneration independently of the others.
A 1-loop diagram has been studied from a similar point of view in \cite{berera}.

\section{Applications}\label{applications}

We conclude with what one might call applications of the preceding results, starting with basic points and then moving on to some thornier questions
involving analytic continuation.  We are everywhere brief and on some points, we simply call attention to questions that might be investigated in the future.

The most basic application was the motivation for this paper:
understanding the Feynman $i\varepsilon$ is crucial for defining loop amplitudes correctly.  

A second immediate point is that this makes unitarity more or less obvious, since unitarity is a general property of 
Feynman diagrams (with real vertices) once the propagators are properly
endowed with the $i\varepsilon$. 
(The traditional understanding of unitarity in string theory has been based on light cone gauge \cite{mandelstam,Giddings,
DPhAoki}.)
String theory amplitudes are similar to field theory amplitudes in this respect -- once the $i\varepsilon$ is in place -- since the singularities in string perturbation theory 
match those that come in field theory from the poles of propagators.   

However, something is still missing.  Instead of borrowing what we know from field theory, one would like to prove unitarity directly in  string perturbation theory.
At the moment it is not clear how to do this.
For example, the  elegant general proof of perturbative unitarity in field theory given in \cite{Diagrammar} uses an equation of greatest time whose
analog in string theory is not obvious.

Another application involves CPT symmetry.  
With a clear $i\varepsilon$ recipe in place, 
it is now even more evident than before that string perturbation theory is CPT-symmetric, just like perturbation theory
in ordinary field theory.  Formally, CPT symmetry in string perturbation theory follows from the fact that the sigma-model describing propagation of a string
is invariant under $X^I\to -X^I$, $I=1,\dots,D$, where the $X^I$ are worldsheet fields describing the motion of the string in $D$-dimensional Minkowski spacetime.  (In asserting that
this is always a symmetry for any Lorentz-invariant string background, we assume that $D$ is even, as is also assumed in the usual statement of the CPT theorem
in field theory.  A slightly modified statement, involving reflection of all coordinates except one of the spatial coordinates,
 holds for odd $D$ in both field theory and string theory.)  Just as in field theory, part of verifying that perturbation theory is CPT-symmetric is CPT symmetry of
the $i\varepsilon$ recipe.  This is so in string theory since the operation $1/L_0\to 1/(L_0-i\varepsilon)$ is CPT-symmetric, just as the operation $1/(p^2+m^2)\to
1/(p^2+m^2-i\varepsilon)$ is CPT-symmetric in field theory.

Another basic question to which the $i\varepsilon$ is relevant is gauge-invariance of perturbative string amplitudes.  To prove gauge-invariance,
one needs to show that a BRST-trivial state $\{Q_B,\W\}$ decouples from an amplitude for scattering of BRST-invariant states.  The proof of this
involves an integration by parts on moduli space (or more exactly on the appropriate integration cycle) and one has to consider
the possibility of surface terms at infinity.  As always, here ``infinity'' refers to possible degenerations of the string worldsheet and  as explained in section
\ref{another}, there are two
essentially different types of degeneration to consider:  generic degenerations in which the momentum 
flowing through the degenerating cycle is generically off-shell, and special degenerations  in which this momentum is
generically on-shell.  Traditionally it is claimed that surface terms can arise only from the special degenerations, and the analysis -- for instance in sections
7 and 8 of \cite{revisited}  -- focuses on those cases.  

The claim that a surface term cannot come from a generic degeneration is usually justified on the following 
grounds.  If the momentum passing through the degenerating cycle obeys a suitable inequality (such that $p^2+m^2>0$ for all relevant string
states, so that the propagator has no singularity) then the integrand in the integral over moduli space
 vanishes at such a degeneration so there can be no surface term.   One then hopes that the 
vanishing in general of the surface term follows from this by analytic continuation.  This argument is a little more roundabout than one might like and is a little
shaky if -- in field theory language -- the momentum passing through the degenerating cycle depends on loop momenta and so is an integration variable (as in fig \ref{offshell}(b)).

The understanding of the Feynman $i\varepsilon$ and the proper integration cycle $\t \Gamma$ for string perturbation theory
makes possible a much sharper argument.  Basically, there is no boundary term at a generic degeneration
because there is no boundary.  The real proper time parameters $\tau$ are integrated up to $\infty$.   Because of  damped oscillatory factors
$\exp(-i\tau(p^2+m^2-i\varepsilon))$ (and negative powers of $\tau$ that come from loop momenta or the period matrix in some special cases; see footnote \ref{log}),
there are no surface terms for $\tau\to\infty$.  The $\tau\to\infty$ behavior in string theory matches the corresponding behavior in field theory because the propagators and
couplings that control the large $\tau$ behavior in string theory can all be imitated by an effective field theory,
and in field theory there are no anomalies in Ward identities from the infrared region $\tau\to\infty$.

Another question that might be more accessible with an understanding of the Feynman $i\varepsilon$ 
is the analytic structure of string scattering amplitudes -- for example, crossing
symmetry, which amounts to 
analytic continuation of two-body elastic scattering from the $s$-channel to the $t$-channel.  Pochhammer 
contours give a complete picture of the analytic properties of tree
amplitudes in string theory, as we reviewed in section \ref{pochhammer}.  However, the analytic properties of loop amplitudes 
are quite complicated (for an introduction to what
happens in field theory, see chapter 16 of \cite{BD}; for a rare example of detailed study of analytic properties of a 1-loop 
amplitude in string theory, see \cite{dph}).  It seems unrealistic
to find an integration cycle that would completely account for the analytic properties of loop amplitudes and what we have 
aimed for in this paper is much less: we have given
a well-defined integral representation of string theory scattering amplitudes when all kinematic invariants (dot products of 
external momenta) are real -- for instance, for real external
momenta in Lorentz signature.  This is precisely what one gets in field theory  from Feynman's  recipe with the $i\varepsilon$.   We have exhibited a string
theory amplitude as an integral of a holomorphic form $\omega$ over a cycle $\t\Gamma$ in a complex manifold $\t\M$.  $\t\Gamma$ is not compact
but (if the kinematic invariants are real) the integral is oscillatory at infinity\footnote{\label{log} This is true in generic regions at infinity.  The exceptions are cases in which the external momenta are
short-circuited in a sense described in section 18.4 of \cite{BD} and do not flow through some of the propagators in a graph.  Convergence of the integral in such a region does not
depend on the external momenta and depends on negative powers of $\tau$ that come from integrals over loop momenta in field theory or from the period matrix
in string theory.  Actually, using a method of chiral splitting that is reviewed in section 4 of \cite{DPhgold}, it is possible to interpret these factors in 
string theory as coming from integration over
loop momenta just as in field theory.  This method is  useful for making the match between field theory and string theory even more transparent.}  and makes sense with the help of a convergence factor $\exp(-\varepsilon\Phi)$.
Though this definition is initially valid only when the kinematic invariants are real, it might be the starting point for analytic continuation in the external
momenta.  One would aim to vary $\t\Gamma$ as the momenta vary so that the integral remains convergent.   This procedure would sometimes fail
and that would lead to singularities of the scattering amplitude.  We will leave this subject for future investigation.  

One last question concerns high energy behavior. String theory scattering amplitudes
are dominated in the high energy, fixed angle regime by a saddle point in the integral over moduli space \cite{gm}.
  In general -- even for the Veneziano amplitude in the physical
regime of positive $s$ or $t$ -- the relevant critical
point can be a complex critical point that is not on the usual real integration cycle.  
It can be tricky in such a situation to understand which critical point dominates the integral.  For an introduction to such matters, see for example
section 2 of \cite{analytic}.  
Such questions may be relevant to a fuller understanding of the asymptotic behavior  of
scattering amplitudes, especially when they are analytically continued from the physical region.

\vskip1cm
 \noindent {\it {Acknowledgements}}  Research supported in part by NSF Grant PHY-0969448. I thank N. Arkani-Hamed, L. Brink, S. Caron-Huot, P. Goddard, and 
 D. Harlow for discussions and S. Mandelstam for pointing out reference \cite{berera}.
 \bibliographystyle{unsrt}

\begin{thebibliography}{99}

\bibitem{gross}
D. J. Gross, ``Superstrings And Unification,'' in {\it 24th International Congress on High Energy Physics}, ed. R. Kotthaus
and J. H. Kuhn (Springer-Verlag, 1989).


\bibitem{berera}
A. Berera, ``Unitary String Amplitudes,'' Nucl. Phys. {\bf B411} (1994) 157-180. 

\bibitem{mandelstam}
S. Mandelstam,  ``Factorization In Dual Models And Functional Integration In String Theory,'' in {\it The Birth Of String Theory}, ed. A. Cappelli et. al. (Cambridge, 2012).



\bibitem{bo}
L. Brink and D. I. Olive, ``Recalculation Of The Unitary Single Planar Dual Loop In The Critical Dimension Of Space Time,''
Nucl. Phys. {\bf B58} (1973) 237-253.


\bibitem{amano}
K. Amano, ``A Finite String Loop Amplitude In A Finite Form,'' Nucl. Phys. {\bf B328} (1989) 510. 


\bibitem{weis}
J. L. Montag and W. I. Weisberger,
``A Finite Representation For A Superstring Scattering Amplitude
And Its Low Energy Limit,'' Nucl. Phys. {\bf B363} (1991) 527-42.

\bibitem{dph}
E. D'Hoker and D. Phong, ``The Box Graph In Superstring Theory,'' Nucl. Phys. {\bf B440} (1995) 24-94. 

\bibitem{Pock}
L. A. Pochhammer, ``Zur theorie der Eulerschen integrale,'' Math. Ann. {\bf 35} 495?-526.

\bibitem{hsh}
A. J. Hansen and J.-P. Sha, ``A Contour Integral Representation For The Dual Five-Point Function And A Symmetry Of The Genus Four Surface in
$\R^6$,'' J. Phys. A: Math. Gen {\bf 39} (2006) 2509-37, arXiv:math-ph/0510064.

\bibitem{KLT}
H. Kawai, S.-H. H. Tye, and D. Lewellyn, ``A Relation Between Tree Amplitudes Of Closed And Open Strings,''
Nucl. Phys. {\bf B269} (1986) 1-23.
 

\bibitem{Mandeltwo}
S. Mandelstam, ``The Interacting-String Picture And Fundamental Interactions,'' in {\it Unified String Theories}, eds. M. B. Green and D. J. Gross (World
Scientific, 1986), 46-102.

\bibitem{Zwiebach}
B. Zwiebach, ``Closed String Field Theory: An Introduction,'' in {\it Gravitation And Quantizations}, eds. J. Zinn-Justin and B. Julia, hep-th/9305026.

\bibitem{wittensurfaces}
E. Witten, ``Notes On Super Riemann Surfaces And Their Moduli,'' arXiv:1209.2459. 

\bibitem{vh}
K. Hori and C. Vafa,  ``Mirror Symmetry,''  hep-th/0002222.

\bibitem{wittenintegrate}
E. Witten, ``Notes on Supermanifolds And Integration,'' arXiv:1209.2199.

\bibitem{more}
E. Witten, ``More On Superstring Perturbation Theory,'' arXiv:1304.2832.

\bibitem{revisited}
E. Witten, ``Superstring Perturbation Theory Revisited,'' arXiv:1209.5461.

\bibitem{sciuto}
 L.~Magnea, S.~Playle, R.~Russo and S.~Sciuto,
  ``Multi-Loop Open String Amplitudes and Their Field Theory Limit,''
  arXiv:1305.6631 [hep-th].

\bibitem{openstring}
E. Witten, ``Noncommutative Geometry And String Theory,'' Nucl. Phys. {\bf B268} (1986) 253. 

\bibitem{gmw}
S. Giddings, E. Martinec, and E. Witten, ``Modular Invariance In String Field Theory,''  Phys. Lett. {\bf B176} (1986) 362.


\bibitem{Giddings}
S. Giddings, ``Conformal Techniques In String Theory And String Field Theory,'' Phys. Reports {\bf 170} (1988) 167-212.

\bibitem{DPhAoki}
K. Aoki, E. D'Hoker, and D. Phong, ``Unitarity Of Closed Superstring Perturbation Theory,'' Nucl. Phys. {\bf B342} (1990)
149-230.

\bibitem{Diagrammar}
G. 't Hooft and M. Veltman, ``Diagrammar,'' in NATO Adv. Study Inst. Ser. B Phys. 4 (1974) 177-322.



\bibitem{BD}
J. D. Bjorken and S. Drell, {\it Relativistic Quantum Fields} (McGraw-Hill, 1965).  

\bibitem{DPhgold}
E. D'Hoker and D. Phong, ``Lectures On Two-Loop Superstrings,'' arXiv:hep-th/0211111.

\bibitem{gm}
D. J. Gross and P. Mende, ``String Theory Beyond The Planck Scale,'' Nucl. Phys. {\bf B303} (1988) 407. 

\bibitem{analytic}
E. Witten, ``Analytic Continuation Of Chern-Simons Theory,'' in {\it Chern-Simons Gauge Theory: 20 Years After}, ed. J. E. Andersen et. al., AMS/IP
Studies in Advanced Mathematics,  arXiv:1001.2933.



\end{thebibliography}

\end{document}